\documentclass[pdflatex,sn-mathphys-num]{sn-jnl}

\usepackage{graphicx}
\usepackage{amsmath,amssymb}
\usepackage{bm}
\usepackage{braket}
\usepackage{xcolor}
\usepackage{tabularx}
\usepackage{hyperref}

\let\origprintabstract\printabstract
\renewcommand{\printabstract}{\newpage\origprintabstract}

\title{Nonlocal Games as Cross-Platform Quantum Benchmarks: Exceeding unconditional classical bounds on trapped-ion processors}

\author[1,5]{\fnm{Anton T.} \sur{Than}}
\author[2]{\fnm{Jim} \sur{Furches}}

\author[3]{\fnm{Debopriyo} \sur{Biswas}}
\author[4,7]{\fnm{Sarah} \sur{Chehade}}
\author[4]{\fnm{Kathleen} \sur{Hamilton}}
\author[3]{\fnm{Bahaa} \sur{Harraz}}
\author[1,5]{\fnm{Xingxin} \sur{Liu}}
\author[3]{\fnm{De} \sur{Luo}}
\author[3]{\fnm{Keqin} \sur{Yan}}
\author[3]{\fnm{Yichao} \sur{Yu}}
\author[3]{\fnm{Vivian Ni} \sur{Zhang}}
\author[3]{\fnm{Liudmila A.} \sur{Zhukas}}

\author[1,5]{\fnm{Alaina M.} \sur{Green}}
\author[3]{\fnm{Alexander} \sur{Kozhanov}}
\author[3]{\fnm{Christopher} \sur{Monroe}}
\author[3]{\fnm{Crystal} \sur{Noel}}

\author*[2,6]{\fnm{Carlos} \sur{Ortiz Marrero}}\email{carlos.ortiz.marrero@colostate.edu, linke@umd.edu}
\author*[1,3,5]{\fnm{Norbert M.} \sur{Linke}}

\affil[1]{\orgdiv{Joint Quantum Institute and Department of Physics}, \orgname{University of Maryland},
  \orgaddress{\city{College Park}, \state{MD 20742}, \country{USA}}}

\affil[2]{\orgdiv{Physical Detection Systems and Deployment Division}, \orgname{Pacific Northwest National Laboratory},
  \orgaddress{\city{Richland}, \state{WA}, \country{USA}}}

\affil[3]{\orgdiv{Duke Quantum Center, Department of Physics, and Department of Electrical and Computer Engineering}, \orgname{Duke University},
  \orgaddress{\city{Durham}, \state{NC 27708}, \country{USA}}}

\affil[4]{\orgdiv{Quantum Information Science Section}, \orgname{Oak Ridge National Laboratory},
\orgaddress{\city{Oak Ridge}, \state{TN}, \country{USA}}}

\affil[5]{\orgdiv{National Quantum Laboratory (QLab)}, \orgname{University of Maryland},
  \orgaddress{\city{College Park}, \state{MD 20742}, \country{USA}}}

\affil[6]{\orgdiv{Department of Computer Science}, \orgname{Colorado State University},
  \orgaddress{\city{Fort Collins}, \state{CO 80524}, \country{USA}}}

\affil[7]{\orgdiv{Quantum Center}, \orgname{University of Tennessee at Chattanooga},
  \orgaddress{\city{Chattanooga}, \state{TN 37403}, \country{USA}}}

\abstract{
Nonlocal games provide application-level benchmarks for quantum hardware whose classical performance bounds are information-theoretic, holding against all classical strategies regardless of computational resources. We implement a 14-vertex graph coloring game, the smallest graph exhibiting a quantum-classical separation for this game type, on four trapped-ion quantum processors across three institutions. One system achieved a win rate that surpasses the classical bound with statistical significance, marking the first violation of a classical bound in a graph coloring nonlocal game on quantum hardware. The remaining systems achieved win rates comparable to the best superconducting processors evaluated on the same game, further illustrating the potential of nonlocal games as cross-architecture quantum benchmarks.
}


\begin{document}

\maketitle

%%% MAIN TEXT %%%

\section{Introduction}

Quantum hardware has advanced significantly in recent years, with platforms such as superconducting qubits, neutral atoms, and trapped ions achieving higher qubit counts and improved gate fidelity~\cite{google2024quantum,reichardt2024fault, ransford2025helios}. As quantum hardware has become larger and more sophisticated, methods of comparing their utility have continued to expand and diversify~\cite{hashim2025practical}. These methods generally fall into three categories: (1) component-level benchmarks - which assess the quality of individual quantum operations on the hardware~\cite{knill2008randomized, helsen2022general,magesan2012efficient, gambetta2012characterization, erhard2019characterizing,nielsen2021gate, blume2017demonstration} - (2) application-level benchmarks - which assess the hardware's ability to successfully harness quantum resources to complete a specific computational task or a suite of such tasks~\cite{tomesh2022supermarq,abughanem2025characterizing, linke2017experimental} - and (3) benchmarks and tests which aim to directly certify the generation and maintenance of the quantum resources which themselves underpin quantum advantage~\cite{friis2018observation, mooney2019entanglement, kam2024characterization,hensen2015loophole,mahadev2022classical,brakerski2021cryptographic}. In this paper, we expand on a benchmarking scheme, previously reported in~\cite{furches2025application}, which combines the advantages of the latter two categories. Specifically, we extend the previous implementation of a nonlocal graph coloring game to additional quantum devices. For the first time, the result from one device tested exceeds the provably optimal classical performance in this game, showing that this proposed benchmark tests not only the ability to execute a computationally relevant task but that its ability to generate and maintain quantum entanglement is critical to its success.

Application-level benchmarks for quantum computers provide a more holistic assessment of a device's \textit{general} computing utility than component-level methods, whose limited power of prediction for quantum algorithm performance has been experimentally demonstrated \cite{linke2017experimental}. However, application-level benchmarks are also limited in their insights. At currently accessible scales, the \textit{general} computing utility of such quantum computations, such as Grover's Search, can only be fairly compared to those of a classical computer limited in allowable gate depth, rather than contemporary classical computers in typical use. Furthermore, whether a device passes or fails the benchmark is somewhat ambiguous as error tolerances vary from demonstration to demonstration. Hence, the ability to gauge how effectively a device can leverage quantum properties for computational advantage is limited. In this work, we expand and demonstrate the utility of benchmarks not for \textit{general} computing ability, but specifically for \textit{quantum} computing ability. An ideal application-level benchmark would certify genuinely quantum behavior through a provable classical performance bound, while simultaneously testing a computationally meaningful task executable on near-term hardware~\cite{proctor2025benchmarking, furches2025application}. Nonlocal games (NLGs) satisfy both requirements. First, NLGs possess provable classical performance bounds: the maximum win rate achievable by any classical strategy is rigorously bounded, so a quantum device that exceeds this bound certifies behavior no classical system can replicate. Second, NLGs define concrete computational tasks (cooperative games between separated players) that can be compiled into shallow circuits executable on current hardware~\cite{furches2025application}. By exceeding a classical bound, a quantum computer unambiguously passes the benchmark, placing NLGs alongside tests that directly certify quantum resources, such as multipartite entanglement generation~\cite{friis2018observation, mooney2019entanglement, kam2024characterization}, loophole-free Bell tests~\cite{hensen2015loophole}, interactive verification protocols~\cite{mahadev2022classical}, and cryptographic tests of quantumness~\cite{brakerski2021cryptographic}, while also assessing the device's computational ability by algorithmically processing quantum information.

Originally derived to study quantum nonlocality, NLGs leverage deep connections to quantum information theory, computational complexity, and operator algebras~\cite{ji2021mip}. NLGs are interactive protocols in which spatially separated parties collaborate to maximize a game-winning probability~\cite{clauser1969proposed, cleve2004consequences}. Crucially, the classical performance bounds in NLGs are information-theoretic: they hold against \emph{all} classical strategies regardless of computational resources, not merely against gate depth-restricted circuits. When a quantum device exceeds such a bound, it certifies that the device produced correlations no classical system can replicate regardless of computational power, providing a qualitatively different guarantee than the computational speedup for \textit{general} computing utility. Certain NLGs can therefore serve as application-level benchmarks~\cite{furches2025application} with clear, unambiguous quantum performance guarantees. More broadly, Kalai et al.~\cite{kalai2023quantum} showed that \emph{any} nonlocal game with a quantum-classical gap can be compiled into a computational advantage; Natarajan and Zhang~\cite{natarajan2023bounding} subsequently proved that the compiled protocol preserves soundness for the CHSH game, yielding a single-prover BQP verification scheme. These results deepen the connection between NLG violations and provable quantum capability.

The use of nonlocal games as experimental quantum benchmarks has gained momentum in recent years. Daniel et al.~\cite{daniel2022quantum} demonstrated cubic Boolean function (CBF) nonlocal games on a trapped-ion system at the University of Maryland, preparing a 6-qubit cyclic cluster state with fidelity of 60.6\% (66.4\% after error mitigation correction) and obtaining results approaching the classical win-rate threshold, beyond which quantum advantage is established. That work leveraged the optimization of native gate transpilation to the trapped-ion gate set, and state-preparation-and-measurement (SPAM) error mitigation; key experimental techniques that we adopt and extend here. More recently, Drmota et al.~\cite{drmota2025experimental} demonstrated quantum advantage in the odd-cycle game with 26$\sigma$ significance using physically separated trapped-ion nodes connected by photonic links. Hart et al.~\cite{hart2025playing} further demonstrated that NLGs can witness topological phase transitions on a quantum computer, broadening the scope of game-based benchmarks. 

\begin{figure}[t]
\centering
\includegraphics[width=\linewidth]{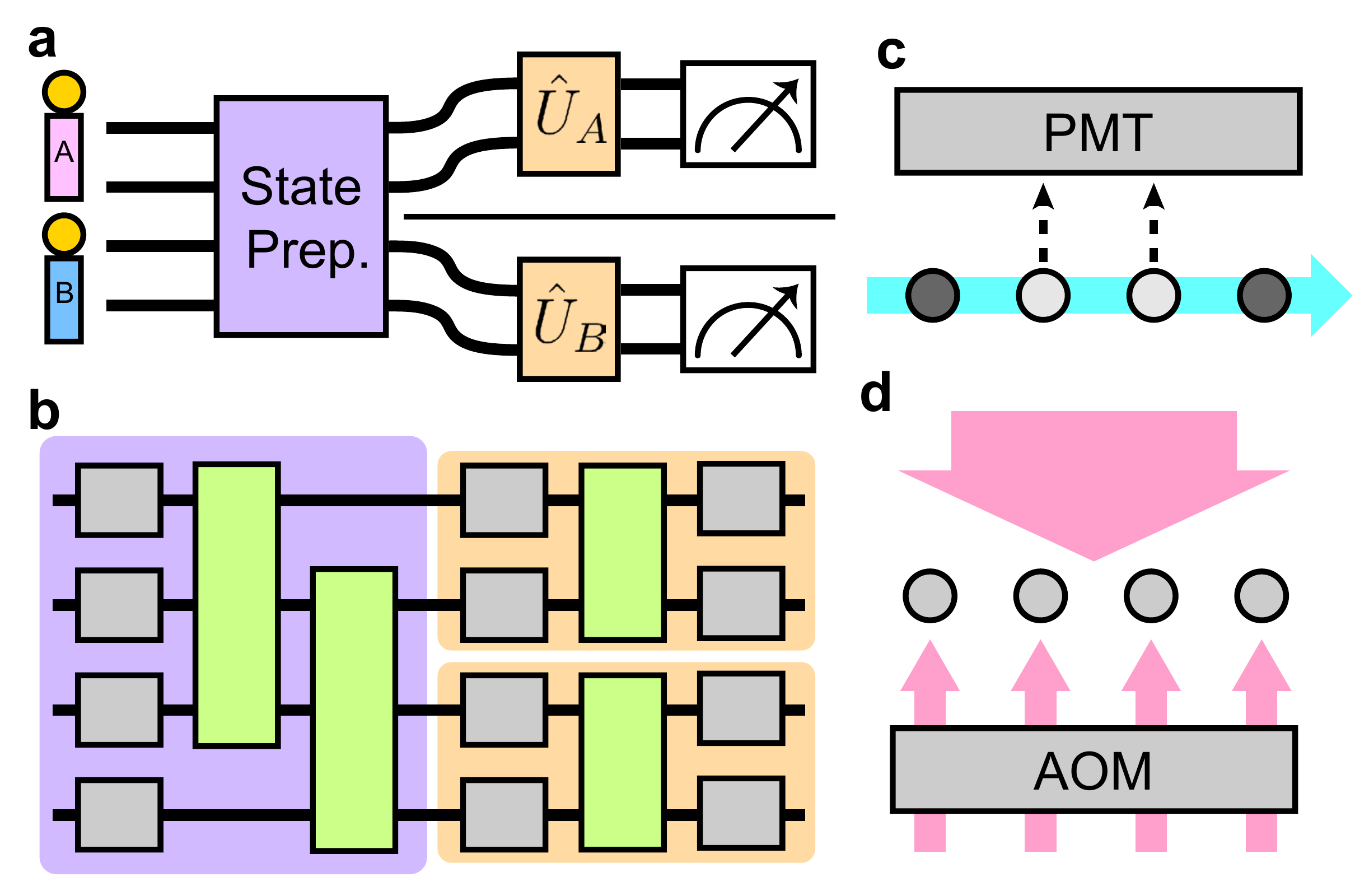}
\caption{(a) Schematic of a nonlocal game. Players A and B each control two qubits. A shared entangled state is prepared (purple, ``State Prep.''), after which the players are separated and each receives a question (a vertex or edge of $G_{14}$) from a referee. Each player applies a question-dependent measurement unitary ($\hat{U}_A$, $\hat{U}_B$) and measures; the two-bit outcomes are interpreted as color assignments, and the referee evaluates whether the coloring rule is satisfied.
(b) Transpiled circuit for the $G_{14}$ game on four qubits. The purple-shaded region prepares two Bell pairs using two M{\o}lmer-S{\o}rensen (MS) gates (green). The orange-shaded region implements the question-dependent measurement unitaries, with one MS gate per player (4 MS gates total per circuit). Gray boxes are single-qubit rotations. One such circuit exists for each of the 51 game questions (14 vertices + 37 edges).
(c) State-dependent fluorescence detection. Resonant laser light (cyan) drives a cycling transition on ions in $\ket{1}$, producing scattered photons collected by a multi-channel photomultiplier tube (PMT). Bright (dark) ions correspond to $\ket{1}$ ($\ket{0}$).
(d) Coherent gate operations via stimulated Raman transitions. A global beam (pink, from above) illuminates all ions, while individual addressing beams pass through a multi-channel acousto-optical modulator (AOM) and are focused onto single ions, enabling independent phase, frequency, and amplitude control for single- and two-qubit gates.
}
\label{fig:main}
\end{figure}

To more firmly establish NLGs as a benchmark uniquely suited to quantum computers, we extend some of the authors' previous application of the $G_{14}$ game to include an additional modality: trapped-ion quantum processors~\cite{furches2025application}. These devices offer long coherence times, high-fidelity two-qubit gates, and flexible qubit connectivity~\cite{debnath2016demonstration}, all of which are beneficial for NLG implementations. In this work, we implement the $G_{14}$ game using four qubits on each of four trapped-ion systems (Fig.~\ref{fig:main}). Among these, the ``Blue System" at Duke University exceeded the classically bounded win rate, demonstrating for the first time a statistically significant quantum advantage in this game. By exceeding the classically bounded win rate of $\omega_c = 86/88 \approx 0.977$, the Blue System's resulting win rate of $\omega = 0.982(3)$ places a milestone on the path to useful quantum advantage, unambiguously surpassing a well-defined threshold. Our approach unifies techniques from theoretical quantum information and experimental benchmarking and continues to support NLGs as a versatile tool for quantum hardware assessment.

The remainder of this paper is organized as follows. After reviewing the theory of nonlocal games and graph coloring games below, Section~\ref{sec:results} presents our experimental results, including win rates, uncertainty analysis, and non-signaling verification. Section~\ref{sec:discussion} discusses the implications for quantum hardware benchmarking, and Section~\ref{sec:methods} details the experimental methods used in this work.

\subsection{General construction}
NLGs are cooperative games. Two or more spatially separated players, who cannot communicate classically once the game begins, provide feedback to an impartial verifier who evaluates their responses based on a predetermined set of rules. A formal description of a two-player NLG with players Alice (A) and Bob (B) is given by $\mathcal{G}=(I_A, I_B, O_A, O_B, \pi, \lambda)$: a finite question set $(I_A, I_B)$, a finite answer set $(O_A,O_B)$, a distribution over the question set $(\pi)$, and a rule function $(\lambda)$. 

To play a round of the game, each player receives a question randomly drawn from the sets $x \in I_A$ and $y \in I_B$ according to the distribution $\pi$. Each player then provides answers $a \in O_A$ and $b \in O_B$ without communicating. The players win this round if the rule function determines whether the answers are valid conditioned on the questions $\lambda(a,b|x,y)=1$.

The maximum probability of winning the game depends on the type of resources available to the players and the strategy they devise. The term \textit{strategy} refers to what resources are shared between the players before they are separated, and how the players implement measurements. Players that rely on \textbf{classical strategies} share classical randomness and deterministic or probabilistic response strategies. 
The \emph{classical win rate} of a nonlocal game is given by:
\begin{equation}\label{eq:classical_game_value}
\omega_c(\mathcal{G}) = \sup_{p(a,b|x,y)} \sum_{x,y} \pi(x,y) \sum_{a,b} \lambda(a,b|x,y) p(a,b|x,y),
\end{equation}
where $p(a,b|x,y)$ obeys local hidden variable models. Players that rely on \textbf{quantum strategies} share a quantum entangled state $\rho$ and perform local measurements $M_x^a$ and $N_y^b$, allowing them to achieve correlations that surpass classical limits. The \emph{quantum win rate} is given by:
\begin{equation}\label{eq:quantum_game_value}
\omega_q(\mathcal{G}) = \sup_{\rho, M_x^a, N_y^b} \sum_{x,y} \pi(x,y) \sum_{a,b} \lambda(a,b|x,y) \text{Tr}\left[(M_x^a \otimes N_y^b) \rho\right].
\end{equation}

A perfect strategy for a NLG is one that wins with probability unity under the ideal verification function, demonstrating that the underlying input-output correlations $p(a,b|x,y)$ exactly satisfy all game constraints. Perfect quantum strategies are of particular interest because their existence certifies that the correlations $p(a,b|x,y)$ required by the game cannot be reproduced by any classical strategy. When a perfect quantum strategy exists but no perfect classical strategy does, the game provides a device-independent witness of nonclassical behavior under the no-communication (non-signaling) assumption.

\subsection{Graph Coloring Games}
Graph coloring games~\cite{paulsen2015quantum} are NLGs where the objective is to find a valid coloring of a graph $G=(V,E)$\cite{harary1969graph}.  A valid coloring maps colors to the vertex set under the constraint that adjacent vertices are never given the same color. If this can be done using $c$ colors, we call this a $c$-coloring of the graph\footnote{A trivial solution can always be found, where a $|V|$-coloring of a graph is found by coloring each vertex a unique color.}. Finding a classical strategy that colors the graph with the fewest colors is equivalent to finding the chromatic number of the graph~\cite{paulsen2015quantum}.

The $G_{14}$ graph coloring game is defined on a 14-vertex graph with an adjacency matrix that challenges classical coloring strategies. For this graph there exists a perfect quantum strategy with $4$ colors, while the smallest possible coloring strategy classically requires $5$~\cite{paulsen2015quantum, mancinska2016oddities}. The best classical strategy using only $4$ colors achieves a win rate of $\omega_c = 86/88 \approx 0.977$, failing on exactly one edge question. $G_{14}$ was discovered in \cite{mancinska2016oddities} and was conjectured to be the smallest possible graph with a quantum advantage for this game. This conjecture was subsequently confirmed in~\cite{lalonde2023quantumchromaticnumberssmall} through exhaustive computational analysis. The perfect quantum strategy is constructed using an orthogonal representation of $G_{14}$ in $\mathbb{R}^4$~\cite{mancinska2016oddities}.

Furches et al.~\cite{furches2025application} use a variational quantum algorithm to generate a perfect (up to numerical precision error) quantum strategy that achieves a quantum chromatic number of $4$ by encoding the rules of the NLG into a Hamiltonian. The resulting quantum strategy serves as a benchmark for quantum computers by testing their ability to implement it. That work executed the strategy on several superconducting quantum hardware platforms; in this work, we extend it to trapped-ion systems.

When playing the graph coloring game to find a valid $c$-coloring, the referee's questions are defined by vertex indices $(x,y)$ and the players Alice and Bob respond with colors $(a,b) \in \{1, \dots, c\}^2$. 
The players execute their strategy through joint measurements on a shared state $\rho$. 
Each is allocated $n=\lceil\log_2{(c)}\rceil$ physical qubits to measure. The answer space $(O_A,O_B)$ is defined by the length-n bitstrings observed by each player after measuring their respective qubits in a vertex-dependent basis.

The questions are drawn from a set that is partitioned into vertex questions and edge questions $(x,y) \sim (I_A, I_B) = \{(v, v) \mid v \in V\} \cup \{(u, v) \mid uv \in E\}$. Vertex questions ask both players to assign a color to the same graph vertex $(v,v)$. If both players observe the same color (bitstring) the answer is valid.  In the $G_{14}$ game there are $4$ valid answers to vertex questions: $|00\rangle |00\rangle$, $|01\rangle|01\rangle$, $|10\rangle |10\rangle$, and $|11\rangle|11\rangle$. 
Edge questions are defined by the undirected graph's adjacency matrix. Both players are given adjacent vertices in the graph $(u, v)$. If both players return different colors (bitstrings) then the answers are valid. In the $G_{14}$ game the remaining  $12/16$ unique outcomes are valid answers to edge questions. 

These constraints are formalized in the rule function $(\lambda)$,
\begin{equation}\label{eq:coloring_rule}
    \lambda(a,b|x,y) = \begin{cases}
        \delta_{a, b} & \text{if } x = y, \\
        (1 - \delta_{a, b}) & \text{if } (x, y) \in E.
    \end{cases}
\end{equation}
The first condition applies to vertex questions and
the second condition applies to edge questions. 
Finally, the players identify an entangled state $\rho$ to share, and the local measurements $M_x^a, N_y^b$.  With the rule function of Eq. \eqref{eq:coloring_rule}, and the choice of $\rho,M_x^a$, and $N_y^b$, the quantum game value of Eq. \eqref{eq:quantum_game_value} can be computed.
\section{Results}
\label{sec:results}
\subsection{Win rate measurements}
\label{subsec:win_rates}

\begin{figure}[t]
\centering
\includegraphics[width=\linewidth]{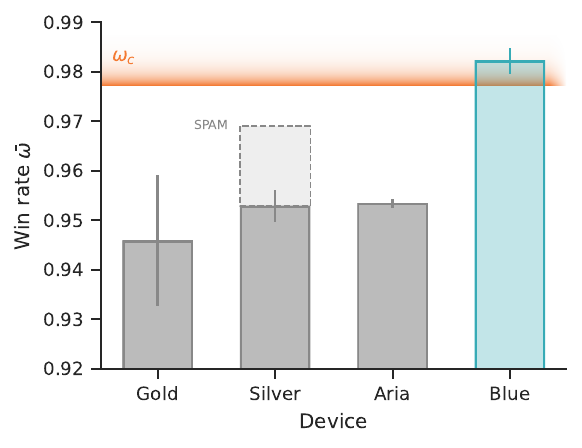}
\caption{\textbf{Win rates for the $G_{14}$ game on four trapped-ion systems.} Win rates are computed using the weighted average in Eq.~\eqref{eq:weighted_winrate}. Error bars denote 95\% confidence intervals computed using concentration inequalities (see Supplementary Note~3). The Blue system achieved $\omega = 0.982(3)$, exceeding the classical limit $\omega_c = 86/88 \approx 0.977$. Solid bars show raw win rates without error mitigation. The dashed bar shows the Silver win rate after SPAM correction via the method in Ref.~\cite{shen2012correcting}, improving from $0.952(3)$ to $0.969$ (see Section~\ref{subsec:error_sources}).}
\label{fig:win_rates}
\end{figure}

Under the uniform question distribution $\pi$, the $G_{14}$ game has 14 vertex questions $(v,v)$ and $37 \times 2 = 74$ directed edge questions $(u,v)$ and $(v,u)$ for each edge $\{u,v\} \in E$. Since each undirected edge circuit covers both directed edge questions (by symmetry of the conjugate measurement protocol), the overall game win rate is
\begin{equation}\label{eq:weighted_winrate}
\omega = \frac{1}{88}\left[\sum_{j \in \mathrm{vertex}} p_j + 2\sum_{j \in \mathrm{edge}} p_j\right],
\end{equation}
where $p_j$ is the win rate of circuit $j$ (estimated from the data as $\hat{p}_j$; see Section~\ref{sec:methods}). The classical game value under this distribution is $\omega_c = 86/88 \approx 0.977$~\cite{furches2025application}: the optimal classical strategy uses a 5-coloring of $G_{14}$, winning all 14 vertex questions deterministically and winning only 72 of the 74 directed edge questions, failing on the single edge whose endpoints share a color in every valid 5-coloring ($1 - \omega_c = 2/88$). We calculated 95\% confidence intervals and significance levels using concentration inequalities (see Supplementary Note~2 for details). The win rates of the devices are shown in Fig.~\ref{fig:win_rates} and Table~\ref{tab:detailed-winrates}. We further plot the win rate for edge and vertex questions separately.

\begin{figure}[t]
\centering
\includegraphics[width=\linewidth]{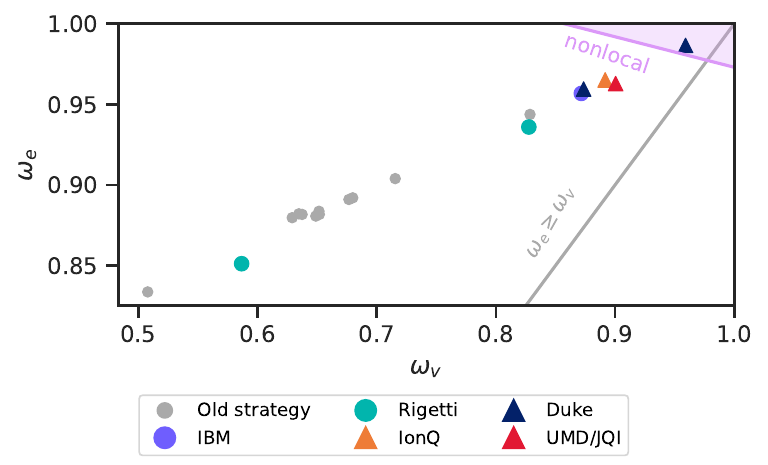}
\caption{\textbf{Comparison of $G_{14}$ benchmarking results across quantum hardware platforms.} Prior experiments on superconducting devices (circles) are from Ref.~\cite{furches2025application}, with colored points indicating comparable circuit depth to this work. Trapped-ion results from this study (triangles) include systems at Duke University (Blue, Gold), the University of Maryland (Silver), and IonQ (Aria). The gray diagonal line marks $\omega_e = \omega_v$; all devices fall above this line, reflecting the asymmetric game rules that make edge questions more noise-robust than vertex questions. The purple curve shows the locus of $(\omega_v, \omega_e)$ pairs satisfying $\omega = \omega_c = 86/88$ under the weighted average in Eq.~\eqref{eq:weighted_winrate}; the shaded region above it is the quantum advantage region. Only the Blue system falls within it. The approximately linear trend across devices reflects an asymmetric sensitivity to noise inherent in the game rules: vertex questions are violated by any single-bit error, whereas edge questions tolerate certain bit flips~\cite{furches2025application}, so both win rates are governed by a common noise parameter and vary in a nearly fixed ratio across platforms.}
\label{fig:frontier_plot}
\end{figure}

The Blue system achieved a win rate of $\omega = 0.982(3)$, surpassing the classical limit $\omega_c$ and demonstrating an advantage consistent with quantum theory ($p = 4.8\times10^{-6}$). The remaining systems (Silver, Gold, and IonQ Aria) did not break the classical threshold, with win rates of $0.952(3)$, $0.945(13)$, and $0.953(1)$, respectively. However, as seen in Fig.~\ref{fig:frontier_plot}, these values are on par with or exceed the best superconducting processors evaluated in~\cite{furches2025application}, indicating that trapped-ion systems performed better in aggregate on this benchmark.

\begin{table}[ht!]
\centering
\begin{tabularx}{\columnwidth}{l|>{\centering\arraybackslash}X>{\centering\arraybackslash}X>{\centering\arraybackslash}X}
\textbf{Device} & $\omega$  & $\omega_v$ & $\omega_e$ \\ \hline
Silver & 0.952(3)  & 0.900      & 0.962      \\
Blue   & 0.982(3)  & 0.959      & 0.986      \\
Gold   & 0.945(13)  & 0.873      & 0.959      \\
Aria   & 0.953(1)  & 0.891      & 0.965
\end{tabularx}
\caption{Game win rates per device and by question type. The overall win rate $\omega$ is computed using Eq.~\eqref{eq:weighted_winrate}, weighting each edge circuit by 2 to match the uniform distribution over 88 game questions. The vertex (edge) win rates are denoted $\omega_v$ ($\omega_e$). Uncertainties for $\omega_v$ and $\omega_e$ are not reported because the Bernstein bound applies to the weighted overall rate $\omega$; the per-question-type rates are shown for completeness.}
\label{tab:detailed-winrates}
\end{table}

Furthermore, similarly to Ref.~\cite{furches2025application}, we observed differences in the vertex and edge question win rates arising from the asymmetric game rules: the edge win rate exceeds the vertex win rate on all devices ($\omega_e > \omega_v$, Table~\ref{tab:detailed-winrates}). This asymmetry has a natural physical explanation: vertex questions require exact bitstring agreement between both players (the ideal probability of matching is $1/4$ per color), while edge questions require only that the players' bitstrings differ, which is a strictly more lenient condition. Hardware noise therefore disproportionately degrades vertex performance. Since edge questions carry twice the weight in Eq.~\eqref{eq:weighted_winrate}, this asymmetry amplifies the overall win rate relative to a naive unweighted average. All devices fall in the region $\omega_e \geq \omega_v$ (Fig.~\ref{fig:frontier_plot}). Notably, both the trapped-ion and superconducting data points cluster along a narrow band in the $(\omega_v, \omega_e)$ plane. This is consistent with a noise model in which hardware errors degrade vertex and edge win rates in a correlated fashion, with vertex questions degrading faster due to their stricter matching requirement.

\subsection{Non-signaling verification}

We note that this experiment does not certify true nonlocality since the qubits of each player reside in close proximity in the ion trap, preventing the spacetime separation required to enforce the no-communication rule. However, we employed the prediction-based ratio (PBR) test of Tabia et al.~\cite{tabia2025almost} to test whether measurement crosstalk could account for the observed win rates (see Supplementary Note~3 for the full methodology). The PBR test computes an upper bound $p_U$ on the p-value for the null hypothesis that the data arise from a non-signaling distribution $P \in \mathcal{NS}$ (the set of all distributions satisfying the no-communication constraint). A small $p_U$ would indicate evidence of signaling, suggesting the high win rate arises from crosstalk rather than genuine quantum correlations. In our analysis, the PBR test returned vacuous bounds of $p_U = 1.0$ for the Blue system data, meaning the test had insufficient statistical power to detect signaling at the levels present in our data. This result is consistent with non-signaling behavior but does not constitute a positive confirmation of it. Achieving a non-trivial bound would likely require substantially more shots or a test statistic with greater sensitivity to the small non-signaling violations that may be present in the data. Moreover, the dominant crosstalk mechanisms in these trapped-ion systems (photon leakage between PMT channels, off-resonant laser scattering; see Section~\ref{subsec:error_sources}) are determined by the physical arrangement of ions and gate parameters, not by the game structure. While these effects may vary across circuits, they degrade state and measurement fidelity rather than inject correlations aligned with the game's winning conditions. This is corroborated by the Silver system data, where correcting for SPAM errors improved the win rate from $0.952(3)$ to $0.969$, confirming that measurement noise degrades rather than enhances game performance.
\subsection{Error analysis and mitigation}
\label{subsec:error_mitigation}

The manual transpilation optimizes the circuit layout, and each circuit requires 4 fully-entangling $R_{xx}$ gates, making this the dominant error source. However, only the Blue system was able to surpass the classical threshold. This is attributed to a two-qubit gate fidelity of $99.5 \pm 0.2\%$ and low measurement crosstalk enabled by fiber-coupled detection with a high-numerical-aperture (NA) objective lens (Supplementary Table~\ref{tab:hardware}).
Meanwhile, SPAM error can be reduced through hardware choices (e.g. the objective lens used in Blue and Gold systems).  It can also be mitigated during post-processing. We used the Silver calibration data to perform SPAM error mitigation as described in~\cite{shen2012correcting}. Without SPAM correction, the Silver system achieved raw win rates of $90.05\%$ for vertex questions and $96.28\%$ for edge questions, yielding an overall weighted win rate of $0.952(3)$ (Eq.~\eqref{eq:weighted_winrate}). After applying SPAM correction, these improved to $93.57\%$ for vertex questions and $97.56\%$ for edge questions, with an overall weighted win rate of $0.969$, still below the classical threshold but significantly improved from the raw measurements. We emphasize that all primary results, including the Blue system's quantum advantage claim, are based on raw (uncorrected) win rates. In the NLG setting, SPAM correction alters the observed outcome distribution and could in principle push a classically achievable win rate above the quantum threshold, undermining the information-theoretic guarantee that distinguishes NLGs from other benchmarks. The Silver SPAM-corrected value is reported only to illustrate the impact of readout error on performance, not as an alternative claim of advantage. Notably, the Blue system achieved quantum advantage without requiring any such correction, demonstrating its superior measurement fidelity.

\section{Discussion}
\label{sec:discussion}

We have demonstrated the $G_{14}$ graph coloring NLG using four ions on each of four trapped-ion quantum processors, achieving the first statistically significant violation of this bound on the Duke Blue system with a win rate of $\omega = 0.982(3)$, exceeding $\omega_c = 86/88 \approx 0.977$. The non-signaling analysis based on the prediction-based ratio test of Tabia et al.~\cite{tabia2025almost} did not detect statistically significant crosstalk, yielding results consistent with non-signaling behavior. The win rate exceeding the classical bound provides evidence that the Blue system exhibits quantum correlations that no classical strategy can reproduce, under the non-signaling assumption supported by the physical analysis of crosstalk mechanisms (Section~\ref{subsec:error_sources}). Our ability to surpass this fundamental classical limit establishes games of this type as a genuine benchmark for quantum performance.

The cross-platform comparison across four systems highlights the sensitivity of NLG benchmarks to global hardware quality. While the Silver, Gold, and Aria systems achieved win rates comparable to the best superconducting processors evaluated in Ref.~\cite{furches2025application}, only the Blue system crossed the classical threshold. This outcome is attributed to its combination of high two-qubit gate fidelity and low measurement crosstalk enabled by a high-NA objective lens. Notably, SPAM correction improved the Silver system's win rate from $0.952(3)$ to $0.969$, approaching the classical threshold and suggesting that reducing measurement error is key to surpassing the classical bound. Unlike traditional benchmarks such as randomized benchmarking, which average over local error channels, the NLG win rate is sensitive to the interplay of gate errors, state preparation, measurement fidelity, and cross-qubit correlations across the full circuit. This makes NLGs a complementary and physically motivated tool for quantum hardware assessment.

However, we note two important caveats. First, both players' qubits are co-located in the same ion trap, preventing enforcement of spacetime separation; our non-signaling test yielded only vacuous bounds and therefore addresses this limitation only in the sense that no signaling was detected, not as a positive confirmation of non-signaling. Second, while the margin above the classical bound ($\sim$0.005) is modest, we note that the Bernstein bound used for significance testing is conservative (one-sided), that calibration drift during circuit execution would tend to increase variance and thus weaken rather than inflate the significance, and that $\omega_c = 86/88$ is an exact value (not an estimate subject to uncertainty). Third, the Aria system was accessed through a cloud platform without independent hardware characterization, so its performance reflects the calibration state at the time of submission rather than a controlled experimental session. Nonetheless, improvements in hardware calibration and shot statistics could strengthen the significance of future demonstrations.

Our work complements that of Drmota et al.~\cite{drmota2025experimental}: our co-located setup trades loophole-free guarantees for cross-platform benchmarking across four systems, enabling hardware ranking that a single-system demonstration does not address.

A key advantage of the graph coloring NLG framework is its natural ability to scale in difficulty. The $G_{14}$ game, with its classical bound of $\omega_c \approx 0.977$, currently discriminates only the highest-fidelity devices; as more systems surpass this threshold, the game will lose its power to rank hardware. However, larger graphs with more vertices and edges require more qubits and deeper circuits, and their classical bounds can be pushed closer to unity, demanding correspondingly higher device fidelity to demonstrate quantum advantage. Different game families, such as the cubic Boolean function games~\cite{daniel2022quantum}, offer complementary scaling along circuit depth rather than qubit count. This provides a mechanism for systematically raising the bar: each generation of quantum hardware can be challenged with a game calibrated to its expected capabilities, ensuring that the benchmark remains informative rather than saturated. In this way, NLG benchmarks can evolve in lockstep with hardware progress, fulfilling the role that progressively harder challenge problems have played in computing.

Looking ahead, several other directions are promising. Connecting remote ion traps via photonic links, as recently demonstrated by Main et al.~\cite{main2025distributed, gottesman1999demonstrating}, could enable true spacetime-separated NLG demonstrations, transitioning from the benchmarking paradigm to loophole-free tests of quantum nonlocality. Alternatively, the no-communication constraint could be enforced computationally rather than physically, using cryptographic barriers that prevent classical simulation of the players' responses~\cite{kalai2023quantum, natarajan2023bounding}. The natural multi-party structure of NLGs also makes them well-suited for certifying the performance of distributed quantum systems and quantum networks~\cite{luo2019nonlocal}. As quantum processors continue to improve, NLGs offer a rigorous, application-level framework for tracking progress toward quantum advantage in a manner grounded in fundamental physics.

\section{Methods}
\label{sec:methods}
\subsection{Hardware description}

The circuits were run on four trapped-ion quantum processors: the Silver system at the University of Maryland, College Park~\cite{debnath2016demonstration}, the Blue and Gold systems at the Duke Quantum Center~\cite{kozhanov2023gold}, and the Aria system at IonQ~\cite{ionq_aria_system}. All four systems encode the qubit in the ground hyperfine clock transition of $^{171}\mathrm{Yb}^+$, with $\ket{0} \equiv \ket{F=0,m_F=0}$ and $\ket{1} \equiv \ket{F=1,m_F=0}$. At the beginning of each experimental run, the ions are cooled near the motional ground state through Doppler and Raman sideband cooling, and the qubits are prepared in the $\ket{0}$ state through optical pumping. Coherent operations are driven using a Raman transition at 355~nm; single-qubit gates drive the qubit transition at 12.6428~GHz resonantly, and two-qubit gates use a M{\o}lmer-S{\o}rensen interaction~\cite{sorensen1999prl, molmer1999prl}. Measurement is performed using state-dependent fluorescence. On the Silver system, photons are focused onto an array of photomultiplier tubes (PMTs), while the Blue and Gold systems use a multi-core fiber to guide fluorescence from each ion to an individual PMT, substantially reducing measurement crosstalk. Circuits were submitted to IonQ's Aria system through a cloud-based access platform without independent characterization of the hardware; the publicly available specifications reported by IonQ are listed alongside the other systems in Supplementary Table~\ref{tab:hardware}.

\subsection{Circuit implementation}
\label{subsec:circuit_impl}

We obtained the quantum circuits implementing the short-depth ``Bell-pair'' strategy from Ref.~\cite{furches2025application} and manually transpiled them to the native gate set of the ion traps: $\{R_\phi(\theta), R_z(\theta), R_{xx}(\theta)\}$. Transpilation was done to minimize the number of two-qubit gates in each circuit, as they were the limiting factor for circuit fidelity; another pass was then done to minimize the number of single-qubit gates. Each circuit contains 4 fully-entangling $R_{xx}(\pi/4)$ gates: 2 for Bell pair preparation and 1 per player's measurement unitary. While a theoretically perfect strategy exists using 2 entangling gates per measurement (6 total per circuit), the 1-gate measurement ansatz was chosen because it achieves a near-perfect game value ($\omega^* = 1 - 2.6 \times 10^{-8}$) while reducing the dominant source of hardware error (see Supplementary Note~4 for analysis). Additionally, two pairs of the fully-entangling gates were able to be done in parallel; this was done on the Silver system using the method in~\cite{zhu2023pairwise} to cut the circuit execution time roughly in half and reduce errors due to decoherence.
In total, each run consisted of 51 circuits, representing the 14 vertices and 37 edges of the graph. Calibration was performed prior to execution and intermittently during the circuit batch to minimize coherent errors (see Section~\ref{sec:methods} for calibration details).

\subsection{Calibration procedure and gate implementation}

Prior to executing game circuits, we performed calibration to minimize coherent over- and under-rotations. For the Silver, Gold, and Blue systems, M{\o}lmer-S{\o}rensen gates were calibrated by performing a measurement to estimate Bell state entanglement fidelity and adjusting gate power and detuning accordingly. For Silver, single-qubit gates were calibrated by adjusting the time to perform a $R_x({\pi/2})$ gate. For the Gold and Blue, single-qubit gates were calibrated using Ramsey experiments to adjust pulse amplitude and phase.

Gate calibrations were done as necessary to account for beam amplitude variation due to experimental drift. Calibration frequency varied by system: before every circuit on Silver, and before each batch on Gold and Blue systems. Additionally, for Blue, laser amplitudes were calibrated automatically through a triggered auto-calibration routine approximately every 10 minutes, or more frequently as needed to compensate for fluctuations in the addressing beam position.

Two-qubit gate implementation also varied across the systems. The Silver system utilized the parallel entangling gate scheme described in \cite{zhu2023pairwise} to minimize circuit time and set the gate time to approximately 250 $\mu s$. On the Gold system, M{\o}lmer-S{\o}rensen gate solutions were classically computed using discrete robust-FM gates described in \cite{kang2021batch, leung2018robust}. Gate durations were fixed at $250~\mu s$, which was chosen to achieve full entangling with the available laser power. Gate detunings were calibrated daily based on measured radial mode frequencies, and laser amplitudes were calibrated hourly to achieve the intended two-qubit gate angle. On the Blue system, entangling gates were implemented using amplitude-modulated (AM) pulses with durations of approximately 140 $\mu s$~\cite{PhysRevA.97.062325}; the AM pulse profile and motional detuning were computed separately for each ion pair to maximize gate performance. 
\subsection{Error sources}
\label{subsec:error_sources}

The two dominant error sources across all four platforms are readout error and two-qubit gate error.

\textbf{State preparation and measurement (SPAM) error.} The leading causes of SPAM error are off-resonant excitations and measurement crosstalk~\cite{debnath2016demonstration}. During the measurement process, off-resonant excitations to undesired states can cause qubits in the $\ket{0}$ state to be measured as $\ket{1}$ or vice-versa. This effect is highly dependent on the ion species; it is roughly the same across all four systems as they all use $^{171}\text{Yb}^+$. Measurement crosstalk arises when fluorescence from ions in the $\ket{1}$ state spills onto neighboring PMT channels, causing neighboring qubits to be erroneously measured as $\ket{1}$. On the Silver system, this error is mitigated by a high numerical aperture (NA) objective lens. On the Blue and Gold systems, measurement crosstalk is further suppressed by fiber-coupled detection, which routes fluorescence from each ion to an individual PMT channel. Meanwhile, the contribution of state preparation to SPAM is small due to the ability to optically pump to the ground state accurately. We note that much of this error can be accounted for through the method in~\cite{shen2012correcting}, as was done in the Silver system (see Section~\ref{subsec:error_mitigation} for why SPAM correction is not applied to primary results).

\textbf{Two-qubit gate error.} The M{\o}lmer-S{\o}rensen entangling gate uses the ions' shared motion as a bus for entanglement. Errors are mostly caused by motional heating and residual spin-motion entanglement~\cite{webb2018resilient,hughes2025trapped}. Motional heating occurs during circuit execution as electric field noise in the environment drives phonon creation in the ion chain, degrading the fidelity of subsequent gates. Residual spin-motion entanglement at the end of the gate, caused by imperfect ground-state cooling or pulse calibration and calculation errors, also leads to decoherence. Heating effects can be mitigated by minimizing circuit execution time; this was done on the Silver system, which utilized parallel entangling gates to halve the execution time~\cite{zhu2023pairwise}. An additional source of error is gate crosstalk, where the addressing beams that drive a two-qubit gate on a target ion pair also weakly drive unwanted transitions on neighboring spectator ions~\cite{fang2022crosstalk}. This introduces incoherent errors that degrade gate fidelity but do not produce question-dependent correlations that could artificially inflate game win rates.
\subsection{Uncertainty quantification}

Following Ref.~\cite{furches2025application}, we bound the deviation of the empirical win rate from its true value using Bernstein's inequality. For each circuit $j$, the empirical win rate $\hat{p}_j$ is computed over $n_j$ independent shots. The weighted win rate $\bar{\omega} = \sum_j w_j \hat{p}_j$ (Eq.~\eqref{eq:weighted_winrate}) is a sum of independent bounded random variables with variance $\sigma_w^2 = \sum_j w_j^2\, p_j(1-p_j)/n_j$. Bernstein's inequality then gives
\begin{equation}
P(|\bar{\omega} - \omega| \geq \epsilon) \leq 2\exp\left(\frac{-\epsilon^2/2}{\sigma_w^2 + \epsilon / 3n}\right),
\end{equation}
where independence across circuits is justified by a fully-reset ion preparation for each circuit and interleaved calibration (see Supplementary Note~2 for full derivation). For significance testing when $\bar{\omega} > \omega_c$, we use the one-sided tail probability with $\epsilon_c = \bar{\omega} - \omega_c$.

\subsection{Non-signaling test}

A probability distribution $P(a,b|x,y)$, where $x,y$ are the referee's questions and $a,b$ are the players' answers, is non-signaling if marginal distributions are independent of the other player's input. Given observed frequencies, we find the closest non-signaling distribution by minimizing Kullback-Leibler (KL) divergence~\cite{cover2006elements} with non-signaling conditions as linear constraints. The prediction-based ratio test statistic $t = \prod_i R(a_i, b_i, x_i, y_i)$ yields upper bound $p_U = \min(t^{-1}, 1)$ via Markov's inequality. We applied 5-fold cross-validation to the Blue system data.

%%% REQUIRED STATEMENTS %%%

\section*{Data availability}

The experimental data and analysis scripts that support the findings of this study are publicly available at \url{https://github.com/cmortiz/IonNLG}.

\section*{Code availability}

The circuit generation and analysis code is publicly available at \url{https://github.com/cmortiz/IonNLG}.

\backmatter

\section*{Acknowledgments}

Thanks to Eleanor Rieffel for providing valuable feedback. This work is supported by a collaboration between the US DOE and other Agencies. This material is based upon work supported by
the U.S. Department of Energy, Office of Science, National Quantum Information Science Research Centers, Quantum Systems Accelerator (Award No. DE-SCL0000121). JF and COM were funded in part by grants from the US Department of Energy, Office of Science, National Quantum Information Science Research Centers, Co-Design Center for Quantum Advantage under contract number DE-SC0012704. S.C.\ was supported in part by the DOE Advanced Scientific Computing Research (ASCR) Accelerated Research in Quantum Computing (ARQC) Program under field work proposal ERKJ354. K.H.\ was supported by the DOE Advanced Scientific Computing Research (ASCR) Pathfinder Testbed Program under FWP ERKJ418. NML acknowledges partial funding by the the National Science Foundation, Software-Tailored Architecture for Quantum Co-Design (STAQ) Award (PHY-2325080).
Access to the IonQ Aria system was made possible by the National Quantum Laboratory (QLab) at the University of Maryland.

This manuscript has been authored in part by UT-Battelle, LLC, under Contract No. DE-AC0500OR22725 with the U.S. Department of Energy. The United States Government retains and the publisher, by accepting the article for publication, acknowledges that the United States Government retains a non-exclusive, paid-up, irrevocable, world-wide license to publish or reproduce the published form of this manuscript, or allow others to do so, for the United States Government purposes. The Department of Energy will provide public access to these results of federally sponsored research in accordance with the DOE Public Access Plan.

\section*{Author contributions}

A.T.T., J.F., C.O.M., and N.M.L.\ conceived the project. A.T.T.\ and J.F.\ developed the methodology. J.F.\ wrote the circuit generation and analysis software, performed validation, and produced all data figures. A.T.T.\ led the experimental implementation, running circuits on the Silver and Aria systems and curating the resulting data. D.B.\ performed experiments and curated data on the Gold system. B.H., D.L., K.Y., and Y.Y.\ contributed to experimental methodology, data collection, and curation on the Blue and Gold systems. V.N.Z.\ contributed to experiments and data curation on the Gold system. L.A.Z.\ performed experiments and contributed to data curation on the Blue system. X.L.\ contributed to experiments and data curation on the Silver system. S.C.\ contributed to the theoretical framework and validation. K.H.\ contributed to the theoretical framework and validation and acquired funding. A.M.G.\ supervised Silver experiments and contributed to data curation. A.K., C.M., and C.N.\ supervised experiments, provided resources for the Blue and Gold systems, and acquired funding. C.O.M.\ supervised the theoretical development, coordinated the writing effort, performed validation, and acquired funding. N.M.L.\ supervised the Silver experiments, provided resources, and acquired funding. All authors contributed to writing and reviewing the manuscript.
\section*{Competing interests}

C.M. is co-founder and Chief Scientist of IonQ, Inc. and has a personal financial interest in the company. All other authors declare no competing interests.

\bibliography{references}

%%% SUPPLEMENTARY INFORMATION %%%
% Supplementary Information
% Demonstration of a Graph Nonlocal Game on Trapped Ion Quantum Computers
% npj Quantum Information format

\newpage
\setcounter{equation}{0}
\setcounter{figure}{0}
\setcounter{table}{0}
\setcounter{section}{0}
\renewcommand{\theequation}{S\arabic{equation}}
\renewcommand{\thefigure}{S\arabic{figure}}
\renewcommand{\thetable}{S\arabic{table}}
\renewcommand{\thesection}{Supplementary Note \arabic{section}}

\section*{Supplementary Information}

\section{Hardware Specifications}\label{supp:hardware}

\begin{table}[h]
\centering
\begin{tabular}{|l|c|c|c|c|}
\hline
 & \textbf{Silver} & \textbf{Gold} & \textbf{Blue}  & \textbf{Aria} \\ \hline
Calibration freq. & Before every circuit & Before batch & Before batch & N/A \\ \hline
SPAM error & $0.83 \%$ & $0.44 \%$ & $0.37 \%$ & $0.39\%$ \\ \hline
Single-qubit fidelity & $99.98 \pm 0.03 \%$ & $99.95 \pm 0.01 \%$ & $99.97 \pm 0.01 \%$ & $99.95\%$\\ \hline
Two-qubit fidelity & $98.6 \pm 0.2 \%$ & $98.55 \pm 0.19 \%$ & $99.5 \pm 0.2 \%$ & $99.6\%$\\ \hline
Circuit runtime & $\sim$550 $\mu$s & $\sim$2.6 ms & $\sim$1.2 ms & $\sim$3 ms\\ \hline
Shots collected & 2000 per circuit & 300 per circuit & 2000 per circuit & 20000 per circuit\\ \hline
\end{tabular}
\caption{Hardware specifications and experimental parameters for the trapped-ion systems. The details of the Aria system were compiled from the IonQ website at \cite{ionq_aria_system}. Calibration frequency refers to the frequency of calibrating single- and two-qubit gates to account for slow drift in experimental parameters. SPAM or state-preparation-and-measurement error is reported as the average of the probabilities of an ion prepared in $\ket{0}$ ($\ket{1}$) state being measured as $\ket{1}$ ($\ket{0}$), and only considers one qubit. Single-qubit fidelity was estimated in different ways. For Silver, this was done by performing a Ramsey experiment, scanning the phase of the second $\pi/2$ pulse, and fitting the amplitude of the resulting sine wave. For Gold and Blue, repeated numbers of $R_x(\pi/2)$ gates are run on a single ion and the fitted decay rate is used to estimate the loss in fidelity per gate. Two-qubit gate fidelity was estimated using even-state populations and parity scan contrast as described in \cite{ballance2016high}. For Silver, the reported fidelity is averaged over all ion pairs used in the experiment, while for Gold and Blue it is a representative value from a single calibrated ion pair. Results for gate fidelity were corrected for SPAM error. Circuit runtime is the total duration of the gate pulses, excluding time for measurement or state preparation.}
\label{tab:hardware}
\end{table}

\section{Uncertainty Calculations}\label{supp:uncertainty}

We derive confidence intervals and significance levels for the game win rate, adapting the concentration inequality framework of Ref.~\cite{furches2025application} to account for the proper weighting of game questions.

\subsection{Concentration inequality}

Using the weighted win rate $\omega = \sum_j w_j p_j$ defined in Eq.~(2) of the main text, with weights $w_j = 1/88$ for vertex circuits and $w_j = 2/88$ for edge circuits, we derive the concentration bound as follows.

For each circuit $j$, let $\hat{p}_j = \frac{1}{n_j}\sum_{i=1}^{n_j} \lambda_{j,i}$ denote the empirical win rate over $n_j$ shots, where $\lambda_{j,i} \in \{0,1\}$ is the win indicator for shot $i$ of circuit $j$. The empirical game win rate is then
\begin{equation}
\bar{\omega} = \sum_{j=1}^{51} w_j\, \hat{p}_j.
\end{equation}
We assume independence across circuits: outcomes from different circuits are independent because each circuit uses freshly prepared ions, and calibration between circuits mitigates temporal correlations. Within each circuit, the $n_j$ shots are i.i.d.\ Bernoulli trials with success probability $p_j$. Under these assumptions, $\bar{\omega}$ is an unbiased estimator of $\omega = \sum_j w_j p_j$ with variance
\begin{equation}
\sigma_w^2 = \mathrm{Var}(\bar{\omega}) = \sum_{j=1}^{51} w_j^2\, \frac{p_j(1-p_j)}{n_j} = \frac{1}{88^2}\left[\sum_{j \in \mathrm{vertex}} \frac{p_j(1-p_j)}{n_j} + 4\sum_{j \in \mathrm{edge}} \frac{p_j(1-p_j)}{n_j}\right].
\end{equation}
To apply Bernstein's inequality, we group by shot. Since $n_j = n$ is the same for all circuits on a given system, define the per-shot outcome $\omega_i = \sum_j w_j \lambda_{j,i}$ for $i = 1, \ldots, n$, so that $\bar{\omega} = \frac{1}{n}\sum_{i=1}^{n}\omega_i$. Since $\omega_i \in [0, \|w\|_1] = [0,1]$, each summand $\omega_i/n$ is bounded by $C = 1/n$, and Bernstein's inequality gives
\begin{equation}
    P(|\bar{\omega} - \omega| \geq \epsilon) \leq 2\exp\left(\frac{-\epsilon^2/2}{\sigma_w^2 + \epsilon / 3n}\right),
\label{eq:supp_win_rate_tail_prob}
\end{equation}
In our experiments, each $p_j$ is estimated from the data as $\hat{p}_j$.

\subsection{Confidence intervals}

The confidence intervals can be obtained from Eq.~(\ref{eq:supp_win_rate_tail_prob}) by setting the right-hand side equal to $\delta$ and solving for $\epsilon$, giving
\begin{equation}
    |\bar{\omega} - \omega| \leq \frac{2\log(2/\delta)}{3n} + \sigma_w\sqrt{2\log(2/\delta)}
    \label{eq:supp_win_rate_sample_error}
\end{equation}
with probability $1 - \delta$, where $\sigma_w$ already incorporates the shot count through the $p_j(1-p_j)/n_j$ terms. We calculate the 95\% confidence interval on the win rates by setting failure probability $\delta = 0.05$ and the shot count $n_j=\{2000, 300, 2000, 20000\}$ for the Silver, Gold, Blue, and Aria processors, respectively.

\subsection{Significance testing}

In the case that $\bar{\omega} > \omega_c$, we would like to confidently assert that the actual win rate $\omega > \omega_c$. We test the null hypothesis $H_0\!: \omega \le \omega_c$ using the one-sided tail probability, which is Eq.~(\ref{eq:supp_win_rate_tail_prob}) without the leading factor of 2. Letting $\epsilon_c = \bar{\omega} - \omega_c$, the null hypothesis probability is
\begin{equation}
    P(\omega \le \omega_c) \le \exp\left(\frac{-\epsilon_c^2/2}{\sigma_w^2 + \epsilon_c/3n} \right).
\end{equation}
This is what we report as the $p$-value for the Blue system, and we claim advantage using a significance level of $\alpha = 0.05$. We did not calculate the significance for the other experiments as $\bar{\omega}$ did not violate the bound $\omega_c$.

\section{Non-signaling Test}\label{supp:non-signaling}

Since the qubits of both players reside in the same ion trap, we cannot enforce true spacetime separation to guarantee the no-communication condition. To verify that our high win rates do not arise from crosstalk (signaling) between qubits, we employ the prediction-based ratio (PBR) test developed by Tabia et al.~\cite{tabia2025almost}.

A probability distribution $P(a,b|x,y)$ is \emph{non-signaling} if the marginal distributions are independent of the other player's input:
\begin{align}
    P(a|x) &= \sum_b P(a,b|x,y) = \sum_b P(a,b|x,y') \quad \forall y, y', \\
    P(b|y) &= \sum_a P(a,b|x,y) = \sum_a P(a,b|x',y) \quad \forall x, x'.
\end{align}
The set of all such distributions is denoted $\mathcal{NS}$. The objective of the PBR protocol is to test the null hypothesis that the observed data from the quantum processor are consistent with a non-signaling distribution. The test returns an upper bound on the $p$-value, and statistically significant values below a threshold $\alpha$ (here chosen prior to the test to be $\alpha = 0.05$) are indicative of crosstalk between the qubits.

We begin by splitting the experimental data from a processor into training and test datasets. First, the closest non-signaling distribution is fit to the training dataset by minimizing the Kullback-Leibler divergence:
\begin{align}
    P^*_{\mathcal{NS}} &= \arg\min_{P \in \mathcal{NS}} D_{\mathrm{KL}}(\vec{f} \| P) \nonumber \\
        &= \arg\min_{P \in \mathcal{NS}} \sum_{a,b,x,y} f(a,b|x,y) \log \frac{f(a,b|x,y)}{P(a,b|x,y)},
\end{align}
where $\vec{f}(a,b|x,y) = N_{a,b,x,y}/N_{x,y}$ are the normalized observed frequencies of the training dataset. We use CVXPY~\cite{diamond2016cvxpy, agrawal2018rewriting} to solve this convex optimization problem with the non-signaling conditions as linear constraints.

Then, over the test dataset, we compute the test statistic
\begin{equation}
    t = \prod_{i=1}^{N} R(a_i, b_i, x_i, y_i)
\end{equation}
where $R(a_i, b_i, x_i, y_i)$ are the prediction-based ratios dependent on the fitted distribution
\begin{equation}
    R(a,b,x,y) = \frac{f(a,b|x,y)}{P^*_{\mathcal{NS}}(a,b|x,y)}.
\end{equation}
The test statistic gives an upper bound on the p-value $p_U = \min(t^{-1}, 1)$.

We applied the PBR test to the Blue system data using k-fold cross-validation with $k = 5$ to ensure the results did not depend on a particular splitting of the data~\cite{hastie2009elements}. Across all folds, we obtained $p_U = 1.0$ (a vacuous bound), with the null hypothesis not rejected at $\alpha = 0.05$. The KL divergence values were at most $D_{\mathrm{KL}}(\vec{f}||P^*_{\mathcal{NS}}) \leq 5.3\times 10^{-4}$, indicating good agreement between our observed data and the non-signaling distribution. The vacuous bound ($p_U = 1$) means the PBR test had insufficient statistical power to detect signaling at the levels present in our experiment. This result is consistent with the absence of crosstalk but does not positively rule it out. A non-trivial bound would require either substantially more measurement shots or a test statistic with greater sensitivity to the small non-signaling violations that may be present in the data.

\section{Circuit Depth Tradeoff for the $G_{14}$ Strategy}\label{supp:circuit_depth}

The quantum strategy deployed on hardware uses measurement unitaries with a single entangling gate (1-$R_{xx}$), yielding circuits with 4 total entangling gates: 2 for Bell pair preparation and 1 per player's measurement. A theoretically perfect strategy exists but requires 2 entangling gates per measurement unitary (6 total). We chose the shallower circuits because the hardware noise saved by eliminating 2 entangling gates far exceeds the negligible theoretical imperfection of $\sim\!10^{-8}$ introduced by the approximate ansatz. Here we analyze both strategies and quantify this tradeoff.

\subsection{Protocol structure}

The shared state is two Bell pairs: $\ket{\Psi} = \frac{1}{2}\sum_{j=0}^{3} \ket{j}_A \ket{j}_B$, prepared using 2 fully-entangling $R_{xx}(\pi/4)$ gates. For vertex $v$, Alice applies unitary $U(v)$ to her qubits and measures in the computational basis. Bob applies $\overline{U(v)}$ (complex conjugate). This conjugate structure ensures the vertex condition $P(a,b|v,v) = \frac{1}{4}\delta_{ab}$ holds automatically.

For adjacent $u \sim v$, the winning condition $P(a,a|u,v) = 0$ for all $a$ is equivalent to the diagonal of $U(u)^\top U(v)$ vanishing:
\begin{equation}\label{eq:supp_edge}
(U(u)^\top U(v))_{aa} = 0 \quad \forall a \in \{0,1,2,3\}.
\end{equation}

\subsection{The deployed strategy: 1-entangling-gate measurement ansatz}

The circuits executed on hardware use a measurement ansatz with a single entangling gate (1-$R_{xx}$), having 8 parameters per vertex (112 total for $G_{14}$). Combined with Bell pair preparation, each circuit contains 4 entangling gates total.

This ansatz cannot achieve a perfect game value ($\omega^* = 1$) due to two independent obstructions:

\textbf{Expressivity obstruction.} The unique perfect strategy requires measurement unitaries in $\mathrm{SO}(4)$ (see below). By the KAK decomposition~\cite{vidal2004universal}, a generic $\mathrm{SO}(4)$ element requires 2 entangling gates. The quaternion matrices required for $G_{14}$ include elements with all four quaternion components nonzero, which cannot be expressed with a single entangling gate.

\textbf{Constraint count obstruction.} The edge condition~\eqref{eq:supp_edge} imposes $37 \times 4 = 148$ constraints. With only 112 parameters, the system is overdetermined by 36 constraints. For generic constraint functions, such systems have no solution.

Numerical optimization of the 1-entangling-gate ansatz achieves a best game value of $\omega^* = 1 - 2.6 \times 10^{-8}$, representing the fundamental limit of this ansatz. This gap persists regardless of state preparation complexity; the bottleneck is measurement expressivity.

\subsection{The perfect strategy via orthogonal representation}

$G_{14}$ admits an orthogonal representation $\varphi: V(G_{14}) \to S^3$ (the unit 3-sphere in $\mathbb{R}^4$), mapping each vertex to a unit vector in $\mathbb{R}^4$ such that adjacent vertices are mapped to orthogonal vectors~\cite{mancinska2016oddities}. For each unit vector $q = (q_0, q_1, q_2, q_3) \in S^3$, define the quaternion matrix $M_q \in \mathrm{SO}(4)$:
\begin{equation}
M_q = \begin{pmatrix}
q_0 & -q_1 & -q_2 & -q_3 \\
q_1 & q_0 & q_3 & -q_2 \\
q_2 & -q_3 & q_0 & q_1 \\
q_3 & q_2 & -q_1 & q_0
\end{pmatrix}.
\end{equation}

The key property is that for $q, p \in S^3$, the diagonal of $M_q^\top M_p$ equals $(\langle q, p\rangle, \langle q, p\rangle, \langle q, p\rangle, \langle q, p\rangle)$. Therefore, orthogonality $\langle q, p \rangle = 0$ implies $\mathrm{diag}(M_q^\top M_p) = 0$, which is precisely the edge condition~\eqref{eq:supp_edge}.

Setting $U(v) = M_{\varphi(v)}^\top$ for each vertex achieves game value $\omega^* = 1$ exactly. Since $M_{\varphi(v)}^\top \in \mathrm{SO}(4) \subset \mathrm{SU}(4)$, and 2 entangling gates suffice for any $\mathrm{SU}(4)$ element by the KAK decomposition~\cite{vidal2004universal}, each measurement unitary requires a 2-entangling-gate circuit, bringing the total to 6 entangling gates per circuit. Numerical decomposition achieves $\|U(v) - e^{i\theta_v} M_{\varphi(v)}^\top\|_F < 2 \times 10^{-6}$ for all vertices, where global phases $e^{i\theta_v}$ cancel in the measurement probabilities. Unconstrained numerical optimization of the 2-entangling-gate parameters (without knowledge of the quaternion structure) independently recovers the quaternion strategy, confirming it is the unique perfect strategy up to equivalence.

\subsection{Practical justification for the approximate strategy}

The choice between the 1-gate (approximate) and 2-gate (perfect) measurement ansätze is governed by a tradeoff between theoretical game value and hardware noise. The theoretical cost of the approximate strategy is a game value reduction of $\Delta\omega_{\mathrm{theory}} = 2.6 \times 10^{-8}$, negligible compared to the classical--quantum gap of $1 - \omega_c \approx 0.023$.

The hardware cost of the perfect strategy is 2 additional entangling gates per circuit (6 vs.\ 4 total). With two-qubit gate fidelities of $\sim$98--99\% on the systems tested (Supplementary Table~1), each additional $R_{xx}$ gate introduces $\sim$1--2\% error. Two extra gates would therefore reduce the measured win rate by an estimated $\sim$2--4\%, far exceeding the margin $\omega - \omega_c \approx 0.005$ by which the Blue system surpasses the classical bound. In other words, the perfect strategy would likely \emph{prevent} observation of quantum advantage on current hardware due to the accumulated gate noise.

The 1-gate ansatz therefore represents the optimal operating point for current trapped-ion hardware: its theoretical imperfection is negligible ($\sim\!10^{-8}$), while the gate count reduction is critical for crossing the classical threshold.

\section{Benchmarking Comparison}\label{supp:benchmarking}
Traditional characterization tools and NLG benchmarks are complementary aspects of quantum hardware. RB and GST estimate \emph{local} noise parameters under structural assumptions, which enables scalable device calibration and error tracking. In contrast, NLGs evaluate the ability of a processor to realize \emph{global} correlations required by a specific information-theoretic task. This difference highlights several implications. 
\begin{itemize}
    \item Correlated and context-dependent errors, often averaged out in RB, directly affect NLG win rates.
    \item NLG performance can be interpreted without assuming gate-independent noise models.
    \item Exceeding a NLG classical bound certifies nonclassical correlations even when the underlying algorithm is classically simulable. 
\end{itemize}

The comparison Table~\ref{tab:benchmark_comparison} below summarizes these distinctions. The categories reflect typical usage rather than strict limitations. NLGs should not be viewed as a replacement for RB or GST. Rather, they address a different question of whether a device can implement the collective correlations required by a concrete protocol whose classical limit is rigorously known. Effective evaluation of quantum hardware thus benefits from using both classes of benchmarks in tandem. 

\begin{table}[h]
    \centering
    \begin{tabular}{l|c|c}
        \hline
        \textbf{Feature}
        & \textbf{RB / GST}
        & \textbf{Nonlocal Games} \\
        \hline
         Benchmark target$^a$ & Individual gates & Multi-qubit correlations \\
         Primary sensitivity$^b$ & Local noise  & Global correlation structure \\
         Correlated errors$^c$ & Largely averaged out & Explicitly amplified \\
         Noise model assumptions$^d$ & Markovian, gate-independent  & Minimal \\
         Scalability$^e$ & High (RB), Moderate (GST) & High \\
         Algorithm relevance$^f$ & Indirect & Direct \\
         Entanglement sensitivity$^g$ & Limited  & High \\
         Quantum--classical separation$^h$  & Absent & Proven \\
     \hline
    \end{tabular}

    \caption{Comparison between traditional quantum benchmarking methods (randomized benchmarking (RB) and gate set tomography (GST)) and nonlocal-game-based benchmarks.}
    \label{tab:benchmark_comparison}
\end{table}
\vspace{0.5em}
    \begin{flushleft}
    $^a$\textit{Benchmark target}: the aspect of device performance a protocol characterizes. RB targets individual gate fidelities via random Clifford sequences~\cite{knill2008randomized, helsen2022general}; GST reconstructs full gate process matrices~\cite{nielsen2021gate, blume2017demonstration}. NLGs test whether the device produces outcome statistics exceeding a game's classical bound, assessing collective circuit performance.\\[0.3em]
    $^b$\textit{Primary sensitivity}: which noise mechanisms most strongly affect the outcome. RB returns an average error rate per gate, sensitive to local noise~\cite{magesan2012efficient}. NLGs are sensitive to how errors in state preparation, entangling gates, and measurement combine across the full circuit.\\[0.3em]
    $^c$\textit{Correlated errors}: noise whose effect depends on gate context (e.g., neighboring gates, qubit crosstalk, temporal drift). RB symmetrizes such errors by averaging over random sequences~\cite{gambetta2012characterization, erhard2019characterizing}. In NLGs, fixed circuits expose correlated errors directly in the win rate.\\[0.3em]
    $^d$\textit{Noise model assumptions}: structural requirements on the error process for interpretable results. RB assumes Markovian, gate-independent noise~\cite{magesan2012efficient, erhard2019characterizing}; GST assumes time-independent process matrices~\cite{nielsen2021gate}. NLGs require only that successive shots are independent.\\[0.3em]
    $^e$\textit{Scalability}: how resource cost grows with qubit count. RB requires only survival probability measurements. GST parameters grow exponentially per gate. For graph coloring NLGs, the strategy is determined by the graph structure, and per-circuit complexity is fixed at any graph size.\\[0.3em]
    $^f$\textit{Algorithm relevance}: how directly a benchmark predicts computational task performance. RB/GST provide indirect information, as gate fidelities may not predict algorithm success~\cite{rieffel2024assessing}. NLGs directly measure success on a task with a known classical limit.\\[0.3em]
    $^g$\textit{Entanglement sensitivity}: whether the outcome depends on the device's ability to generate entangled states. RB can use product states and single-qubit Cliffords. NLGs require entanglement to exceed the classical bound, directly probing multi-qubit entanglement quality.\\[0.3em]
    $^h$\textit{Quantum-classical separation}: a proven bound on the best classical performance that a quantum device can exceed. RB/GST quantify noise but have no such bound. NLGs possess information-theoretic bounds: exceeding $\omega_c$ certifies correlations no classical strategy can reproduce.
    \end{flushleft}

\end{document}